\documentclass[pra,preprint]{revtex4-1}
\usepackage{graphicx}
\usepackage{epstopdf}
\newcommand{\e}{{\rm e}}

\newcommand{\bv}{{\bf v}}

\newcommand{\x}{{\bf x}}
\newcommand{\bx}{{\bf X}}
\newcommand{\s}{{\bf s}}

\newcommand{\ld}{{\lambda}}

\newcommand{\bea}{\begin{eqnarray}}
\newcommand{\eea}{\end{eqnarray}}
\newcommand{\be}{\begin{equation}}
\newcommand{\ee}{\end{equation}}
\newcommand{\ba}{\begin{eqnarray}}
\newcommand{\ea}{\end{eqnarray}}

\newcommand{\nn}{\nonumber}
\newcommand{\la}{\label}

\begin{document}
\title{Solving fermion problems without solving the sign problem: symmetry-breaking wave functions from similarity-transformed propagators for solving 2D quantum dots}

\author{Siu A. Chin}
\affiliation{Department of Physics and Astronomy,
Texas A\&M University, College Station, TX 77843, USA}

\begin{abstract}

It is well known that the use of the primitive second-order propagator in Path Integral Monte Carlo calculations of many-fermion systems leads to the sign problem. In this work, we show that by using the similarity-transformed Fokker-Planck propagator, it is possible to solve for the ground state of a large quantum dot, with up to 100 polarized electrons, without solving the sign problem. These similarity-transformed propagators naturally produce rotational symmetry-breaking ground state wave functions previously used in the study of quantum dots and quantum Hall effects. However, instead of localizing the electrons at positions which {\it minimize} the potential energy, this derivation shows that they should be located at positions which {\it maximize} the bosonic ground state wave function.
Further improvements in the energy can be obtained by using these as initial wave functions in a Ground State Path-Integral Monte Carlo calculation with second and fourth-order propagators.

\end{abstract}


\maketitle

\section{Introduction}

Two dimensional, circular parabolically confined quantum dots, are not only
physical systems of great experimental interests\cite{rei02}, but are also mathematical
models {\it par excellence} for the numerical study of the many-fermion problem. 
In constrast to real atoms, where the hydrogen atom's partition function is 
divergent\cite{bli95}, these ``Hooke's atoms"\cite{nei62} only have bound states, with convergent partition functions.
This lack of additional complications allows us to focus attention solely on the effect of 
interaction and Fermi statistics. In this work we compute the ground state energy of
up to $N=100$ spin-polarized electrons, applicable to the study of
quantum dots under strong magentic fields.

Quantum dots have been extensively studied by traditional methods of quantum many-body
theory, such as Hartree-Fock (HF)\cite{yan99}, Density Functional Theory (DFT)\cite{fer94,hir99}, 
Configuration Interaction (CI)\cite{ron06}, 
Coupled-Cluster (CC)\cite{ped,wal}, Variational Monte Carlo (VMC)\cite{har02,kai02}, Diffusion Monte Carlo 
(DMC)\cite{ped,ped03,gho07} 
and Path Integral Monte Carlo (PIMC)\cite{mak98,egg99,egg00,reu03,chin15,ilk17}, with varying degrees of accuracy. However, with 
increasing number of electrons (say $N\!>$10), basis-function based methods, such as CI and CC,
simply cannot keep up with the exponential growth of needed basis functions.
For $N>20$, even VMC and DMC have difficulties in constructing a good trial wave function 
involving many excited states. In principle, since PIMC does not require an
initial trial wave function, it can be used to treat large quantum dots. However, PIMC can
only extract the ground state at large imaginary time, and if many short-time anti-symmtric
propagators are used, then the resulting {\it sign-problem} will overwhelm the ground state signal. 
One can side-step the sign problem in DMC and PIMC by invoking the fixed-node or 
the restricted-path approximation\cite{ilk17,cep95}. These approximations have worked surprising 
well and currently the ground state energy of the largest spin-balanced quantum dot with $N=60$ 
has been obtained using PIMC\cite{ilk17}. Here, we proposes a new way of solving the fermion
problem in large quantum dots without invoking any prior assumptions. 

In Ref.\cite{chin15}, it was suggested that fourth-order propagators can be used in PIMC to reduce the number of anti-symmtric propagators used and thereby reduce the serverity of the sign problem. This is indeed 
a workable scheme for up to $N\approx$ 30. However, beyond that, the sign problem remains severe at
large imaginary time.

In this work, we overcome this fundamental problem by reducing the length of the imaginary time
needed by doing PIMC on symmetry-breaking wave functions that are already very close to the ground state,
that is, we apply a Fermion Ground State Path Integral Monte Carlo (FGSPIMC) method
to quantum dots. While the bosonic GSPIMC method is well known\cite{cep95,sar00},
the fermionic version has only been tried previously in the context of shadow wave function\cite{cal14}.

To derive such a symmetry-breaking wave function, we first derive, from a new perspective, some basic results on similarity transformed propagators in Section \ref{stp}. In Section \ref{thp}, we show that the harmonic oscillator has the remarkable property that if its propagator is similarity-transformed by its ground state wave function, the resulting Fokker-Planck propagator, even if only approximated to first order, yields the {\it exact} partition function of the harmonic oscillator. We show in Section \ref{non} that, when these Fokker-Planck propagators are anti-symmetrized in the many-fermion case, they yielded the {\it exact} ground state energies of non-interacting fermions in a harmonic oscillator. That is, a many-fermion problem has been solved exactly without knowing the exact propagator, the exact wave functions, or having to solve 
any sign problem.  In Section \ref{sbwf}, we show that in the presence of pair-wise repulsive Coulomb interactions, the resulting Fokker-Planck propagator naturally produces spontaneous symmetry-breaking (SSB) wave functions previously used in the studies of quantum dots and quantum Hall effects\cite{yan99,mik01,yan02,yan07}. For quantum dots, we show that a variational version of these SSB wave functions can already yield energies to within 1$\%$ of the best ground state energies. In Section \ref{gspimc}, we show that this remaining 1$\%$ can be recovered by doing a FGSPIMC calculation using a fourth-order propagator. In Section \ref{con}, we summarize our conclusions and suggest furture applications of this work.       

\section{Similarity transformed propagators}
\la{stp}

For completeness, we will derive here some basic results in a new way.
Let the imaginary-time propagator (or density matrix) of the Hamiltonian operator $H$ be
\be
G(\x,\x';\tau)=\langle \x|\e^{-\tau H}|\x'\rangle,
\la{propa}
\ee
then corresponding partition function
\be
Z(\tau)=\int d\x\,  G(\x,\x;\tau)
\la{gh}
\ee 
is invariant under the similarity transformation
$$
Z(\tau)=\int d\x \langle \x|\phi \e^{-\tau H}\phi^{-1}|\x\rangle
=\int d\x \phi(\x)G(\x,\x;\tau)\phi^{-1}(\x)=\int d\x\,  G(\x,\x;\tau),
$$ 
provided that $\phi(\x)$ is a non-vanishing real function at all $\x$,
\be
\phi(\x)\ne 0.
\la{nonvan}
\ee
Therefore, $Z(\tau)$ can also be computed from the transformed propagator
$$
\tilde G(\x,\x';\tau)=\langle \x|\e^{-\tau \tilde H}|\x'\rangle
$$
corresponding to the transformed Hamiltonian
\be
\tilde H=\phi(\x)H\phi^{-1}(\x).
\la{hp}
\ee
Since $\phi(\x)$ is non-vanishing everywhere, it can always be written as
\be
\phi(\x)=e^{-S(\x)},
\ee
which defines $S(\x)$. We will call $S(\x)$ the {\it action} of the wave function.
For a single-particle Hamiltonian in $D$-dimension of the separable form,
$$
H=-\frac12\nabla^2+V(\x)=K+V(\x),
$$
the transformed Hamiltonian is
$$
\tilde H=\e^{-S(\x)}K\e^{S(\x)}+V(\x).
$$
Since $K$ is only a second-order derivative operator,
the general operator identity
\be
\e^CK\e^{-C}=K+[C,K]+\frac1{2!} [C,[C,K]]+\frac1{3!}[C,[C,[C,K]]]+\cdots,
\la{id}
\ee
with $C=-S(x)$, terminates at the double commutator term:
$$
\e^{-S(\x)}K\e^{S(\x)}=K-[S,K]+\frac1{2} [S,[S,K]].
$$
From the definition of $K=-\frac12\nabla^2$, one has
$$
[S,K]=\nabla S\cdot\nabla +\frac12 \nabla^2 S,\qquad [S,[S,K]=-(\nabla S)^2
$$
and therefore the transformed Hamiltonian is
\ba
\tilde H\psi
&=&(K+D+E_L)\psi
\nn
\ea
where $D$ is the {\it drift} operator
$$
D\psi=\nabla\cdot\Bigl(\bv(\x) \psi \Bigr)
$$
with drift velocity $\bv(\x)=-\nabla S(\x)$ and
$$
E_L(\x)=\frac12 \nabla^2 S-\frac12(\nabla S)^2+V=\frac{H\phi(\x)}{\phi(\x)}
$$
is the {\it local energy} function. The transformed imaginary time propagator is then
\be
\tilde G(\x,\x';\tau)=\langle \x|\e^{-\tau (K+D+E_L) }|\x'\rangle.
\la{ghp}
\ee
The present derivation of this fundamental result on the basis of (\ref{id})
is new, as far as the author can tell. 
 
If $E_L$ is a constant, then because of the non-vanishing condition (\ref{nonvan}), $\phi(\x)$ must be 
the {\it bosonic} ground state $\psi_0(\x)$ of $H$. In this case, (\ref{ghp}) is the {\it Fokker-Planck} (FP)
propagator whose long time stationary solution is the square of the ground state wave function:
$\phi^2(\x)=\psi_0^2(\x)$. 

Even in cases where $E_L$ is not a constant, the advantage of using the transformed propagator (\ref{ghp}) 
is that low order approximates of $\tilde G(\x,\x';\tau)$ can
be far more accurate than low order approximates of $G(\x,\x';\tau)$. For example,
a first-order (in $\tau$) approximation of (\ref{ghp}) is
\be
\tilde G_1(\x,\x';\tau)=\langle \x|\e^{-\tau K}\e^{-\tau D }|\x'\rangle\e^{-\tau E_L(\x)}.
\ee
Since as shown in Ref.\cite{chin90}, 
\ba
\langle \x|\e^{-\tau K }|\x_1\rangle&=&(2\pi\tau)^{-D/2}
\exp\left[-\frac1{2\tau}(\x-\x_1)^2\right],\nn\\
\langle \x_1|\e^{-\tau D}|\x_0\rangle&=&\delta[\x_1-\x(\tau)],
\nn
\ea
where $\x(\tau)$ is the solution to the drift equation with initial position $\x_0$:
$$
\frac{d\x}{d\tau}=\bv(\x)=-\nabla S(\x),
$$
the resulting first-order propagator is
\ba
\tilde G_1(\x,\x_0;\tau)&=&\int d\x_1\langle \x|\e^{-\tau K }|\x_1\rangle
\langle \x_1|\e^{-\tau D}|\x_0\rangle\e^{-\tau E_L(\x_0)}\nn\\
&=&\frac{1}{(2\pi\tau)^{D/2}}
\exp\left[-\frac1{2\tau}(\x-\x(\tau))^2\right]\e^{-\tau E_L(\x_0)}.
\la{gp1}
\ea
This is to be compared with the first-order approximation of $G(\x,\x_0;\tau)$:
\ba
G_1(\x,\x_0;\tau)&=&\int d\x_1\langle \x|\e^{-\tau K }\e^{-\tau V }|\x_0\rangle\nn\\
&=&\frac{1}{(2\pi\tau)^{D/2}}
\exp\left[-\frac1{2\tau}(\x-\x_0)^2\right]\e^{-\tau V(\x_0)}.
\la{g1}
\ea
The transformed propagator (\ref{gp1}) resplaces the bare potential $V(\x)$, which can be highly singular, by $E_L(\x)$, which
can be a non-singular and less fluctuating. It also replaces the aimless Gaussian random walk in $G_1(\x,\x_0;\tau)$ by Gaussian random walks along trajectories of the velocity field $\bv(\x)=-\nabla S(\x)$ produced by the trial wave function. 
This transformed propagator $\tilde G_1(\x,\x_0;\tau)$ is the basis for doing DMC\cite{mos82,rey82} with importance-sampling and is the generalized Feynman-Kac path integral\cite{caf88} when $\phi(\x)\ne\psi_0(\x)$.
In the next section, we will show that this FP propagator produces a remarkable result for the
harmonic oscillator.

\section{Transformed harmonic propagators}
\la{thp}

Consider a $D$-dimensional harmonic Hamiltonian with energy in units of $\hbar\omega$ 
and length in units of $\sqrt{\hbar/m\omega}$,
$$
H=-\frac12\nabla^2+\frac12 \x^2.
$$
In this case, one can take $\phi(\x)=\psi_0(\x)$, the exact ground state wave function with
action
$$
S(\x)=\frac12 \x^2,
$$
and a constant $E_L$,
\ba
E_L&=&\frac12\nabla^2 S-\frac12(\nabla S)^2
+\frac12 \x^2 =\frac{D}2\equiv E_0,
\ea
 which is the exact ground state energy.

The solution $\x(\tau)$ to the drift equation with initial position $\x_0$ is then simply
$$
\frac{d\x}{d\tau}=\bv(\x)=-\nabla S(\x)=-\x\quad\rightarrow\quad \x(\tau)=\x_0\e^{-\tau},
$$
giving the first-order transformed propagator (\ref{gp1}):
\ba
\tilde G_1(\x,\x_0;\tau)
&=&\frac{1}{(2\pi\tau)^{D/2}}
\exp\left[-\frac1{2\tau}(\x-\x_0\e^{-\tau} )^2\right]\e^{-\tau E_L}.
\la{tph1}
\ea
This is to be compared to the exact FK propagator, corresponding to the
Ornstein-Uhlenbeck\cite{uhl} process:
\be
\tilde G(\x,\x_0;\tau)=\frac{1}{[2\pi T(\tau)]^{D/2}}
\exp\left[-\frac1{2T(\tau)}(\x-\x_0\e^{-\tau} )^2\right]\e^{-\tau E_L}
\la{fph}
\ee
with 
\be
T(\tau)=\frac12 (1-\e^{-2\tau}).
\la{ext}
\ee
In the limit of $\tau\rightarrow\infty$, this exact FK propagator correctly gives
\be
\tilde G(\x,\x_0;\tau)\rightarrow\frac{1}{\pi^{D/2}}
\exp\left[-\x^2\right]\e^{-\tau E_0}=\psi^2_0(\x) \e^{-\tau E_0},
\la{exwf}
\ee
which is proportional to the {\it square} of the ground state wave function.
By contrast, the first-order transformed propagator $\tilde G_1(\x,\x_0;\tau)\rightarrow 0$ as $\tau\rightarrow\infty$ and bears no resemblance to any wave function. 
This seems to be a very poor approximation to the exact propagator.
However, if one computes the partition function from this {\it single} transformed propagator,
\ba
Z&=&\int d\x \tilde G_1(\x,\x;\tau)=\frac1{(2\pi\tau)^{D/2}}\int d\x 
\exp\left[-\frac1{2\tau}\x^2(1-\e^{-\tau} )^2\right]\e^{-\tau E_L}\la{zg}\\
&=&\frac1{(2\pi\tau)^{D/2}}[2\pi\tau(1-\e^{-\tau} )^{-2}]^{D/2}\e^{-\tau D/2}
=\left( \frac{\e^{-\frac12 \tau}}{1-\e^{-\tau}} \right)^D
=[2\sinh(\tau/2)]^{-D},
\ea
the result is {\it exactly} correct. That is, when the exact ground state wave function,  
which knows nothing about $\tau$, is used to derived the transformed propagator, 
the resulting single bead calculation produces the correct $Z(\tau)$ at all $\tau$,
{\it i.e.}, at all temperature!

The only difference between the transformed first-order propagator (\ref{tph1}) and the
exact FK propagator (\ref{fph}) is that the variance of the Gaussian distribution is $\tau$
rather than $T(\tau)$. This single bead calculation of $Z(\tau)$ is exact because the
variance of the Gaussian distribution, after doing the integral, is cancelled by the initial 
normalization constant and the integral is actually independent of the variance. 
This suggests that the solution to the drift equation,  which is purely classical, 
is of unexpected importance for understanding
quantum statistical dynamics, at least for the harmonic oscillator. 
In the next Section, we will see how the drift term exactly solves the
problem of many non-interacting fermions in a harmonic oscillator, without knowing the 
exact harmonic oscillator propagator.

\section{Non-interacting fermions in a harmonic oscillator}
\la{non}

In (\ref{exwf}), one sees that the exact FP propagator yields the {\it square} of the
ground state wave function in the limite of $\tau\rightarrow\infty$ with
$$
\x(\tau)\rightarrow 0\quad{\rm and}\quad T(\tau)\rightarrow \frac12.
$$
In the first-order transformed propagator (\ref{tph1}), one also has $\x(\tau)\rightarrow 0$
as $\tau\rightarrow\infty$. What is left is then a Gaussian distribution with variance $\tau$.
If one now regards this variance $\tau$ as just a variational parameter, and dissociate it
from being the imaginary time needed to be set to infinity,
then the choice of $\tau=1$ would give the correct ground state wave function (but not
its square). This seems to be a rather contrived way of obtaining the ground state wave function
from the transformed propagator, but its utility is the following. 

\begin{figure}[hbt]
\includegraphics[width=0.49\linewidth]{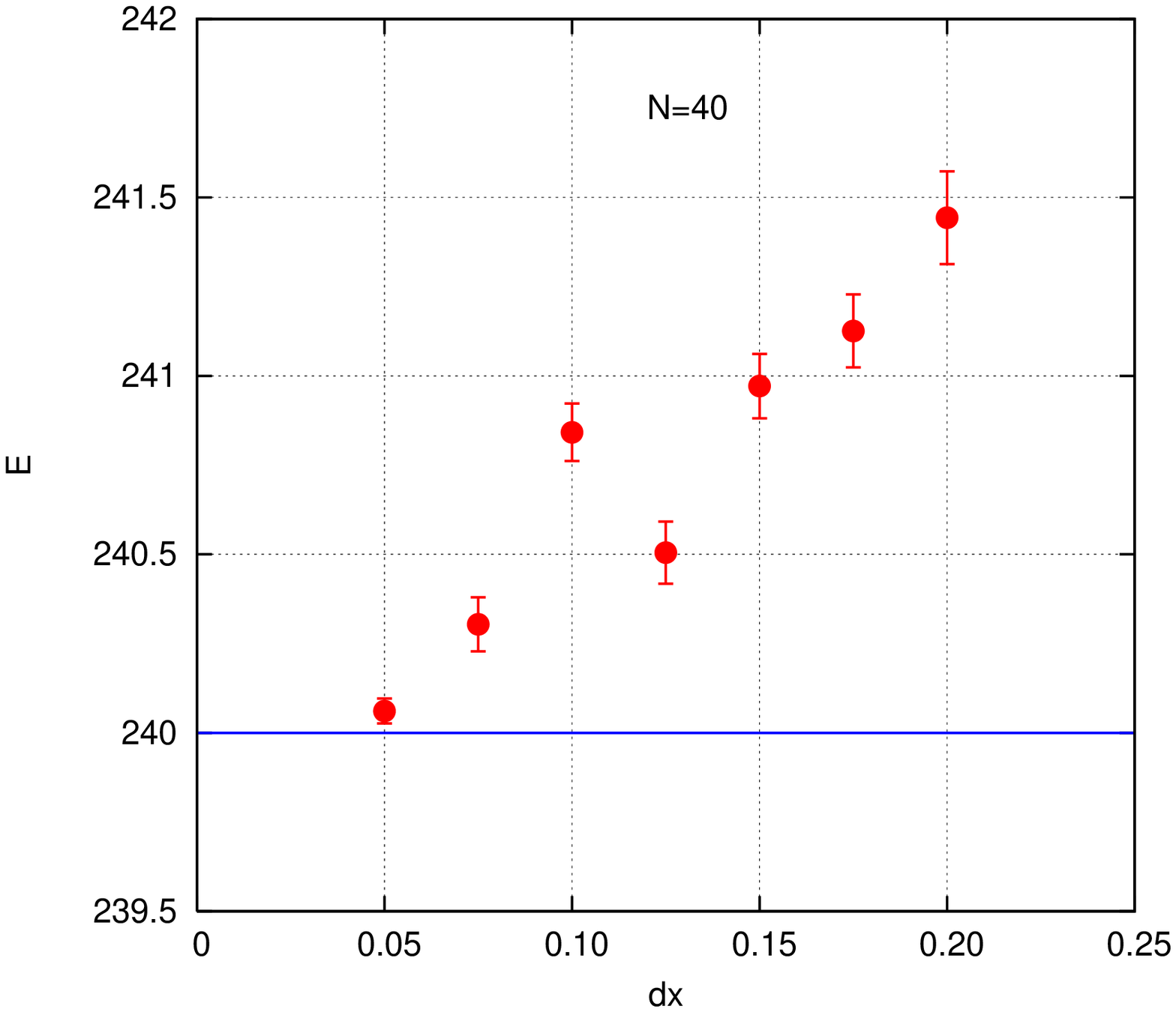}
\includegraphics[width=0.49\linewidth]{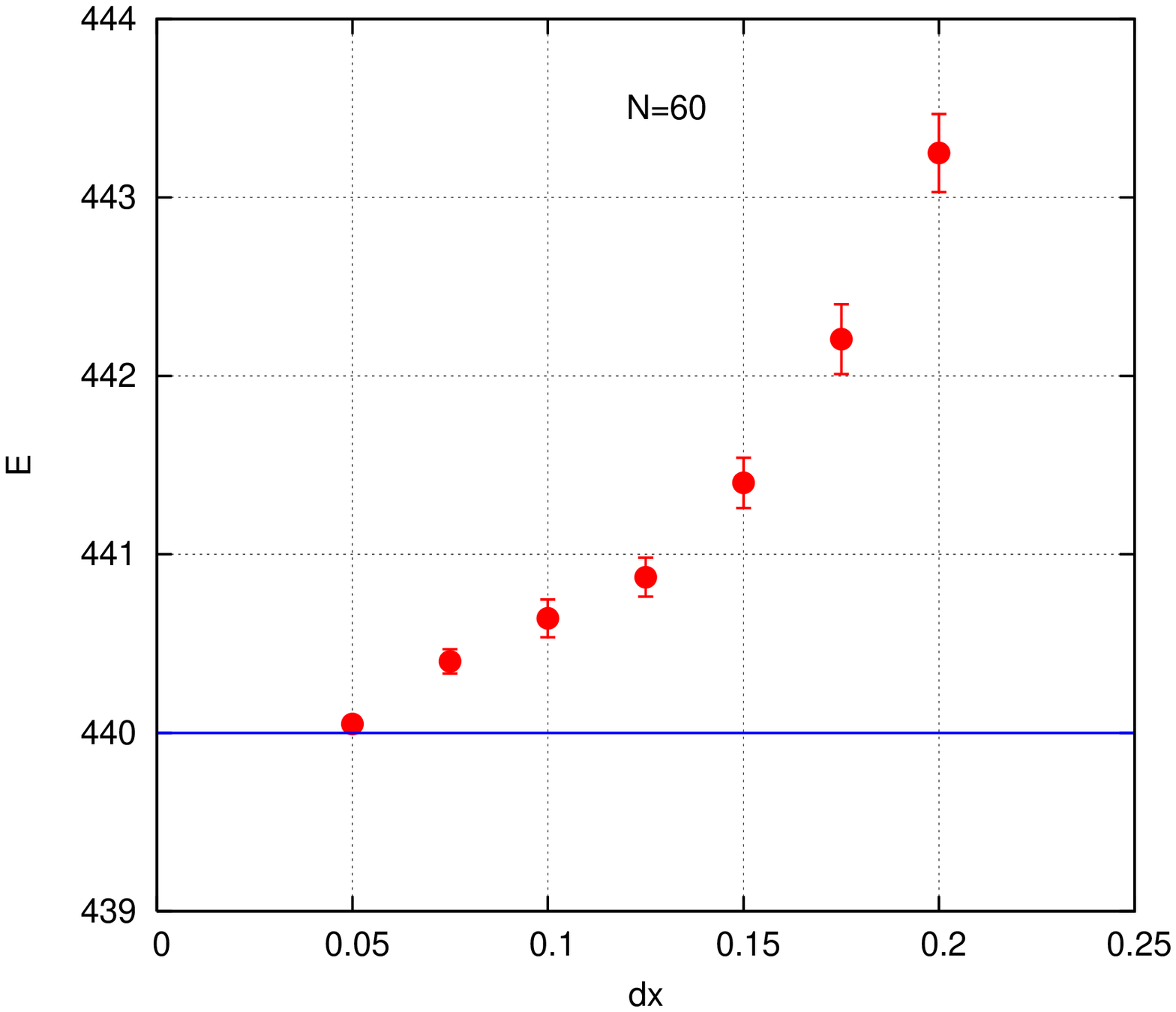}

\includegraphics[width=0.49\linewidth]{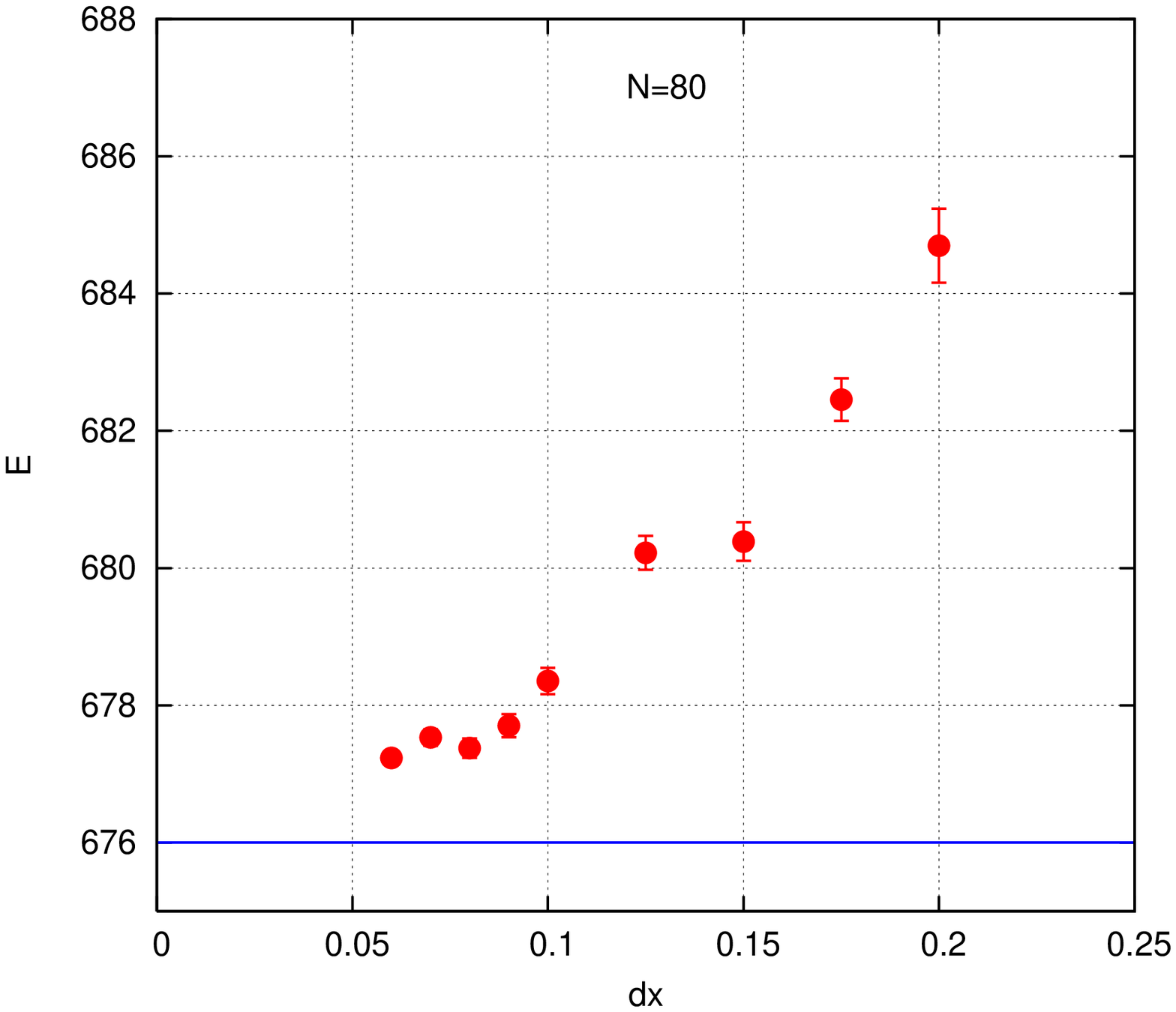}
\includegraphics[width=0.49\linewidth]{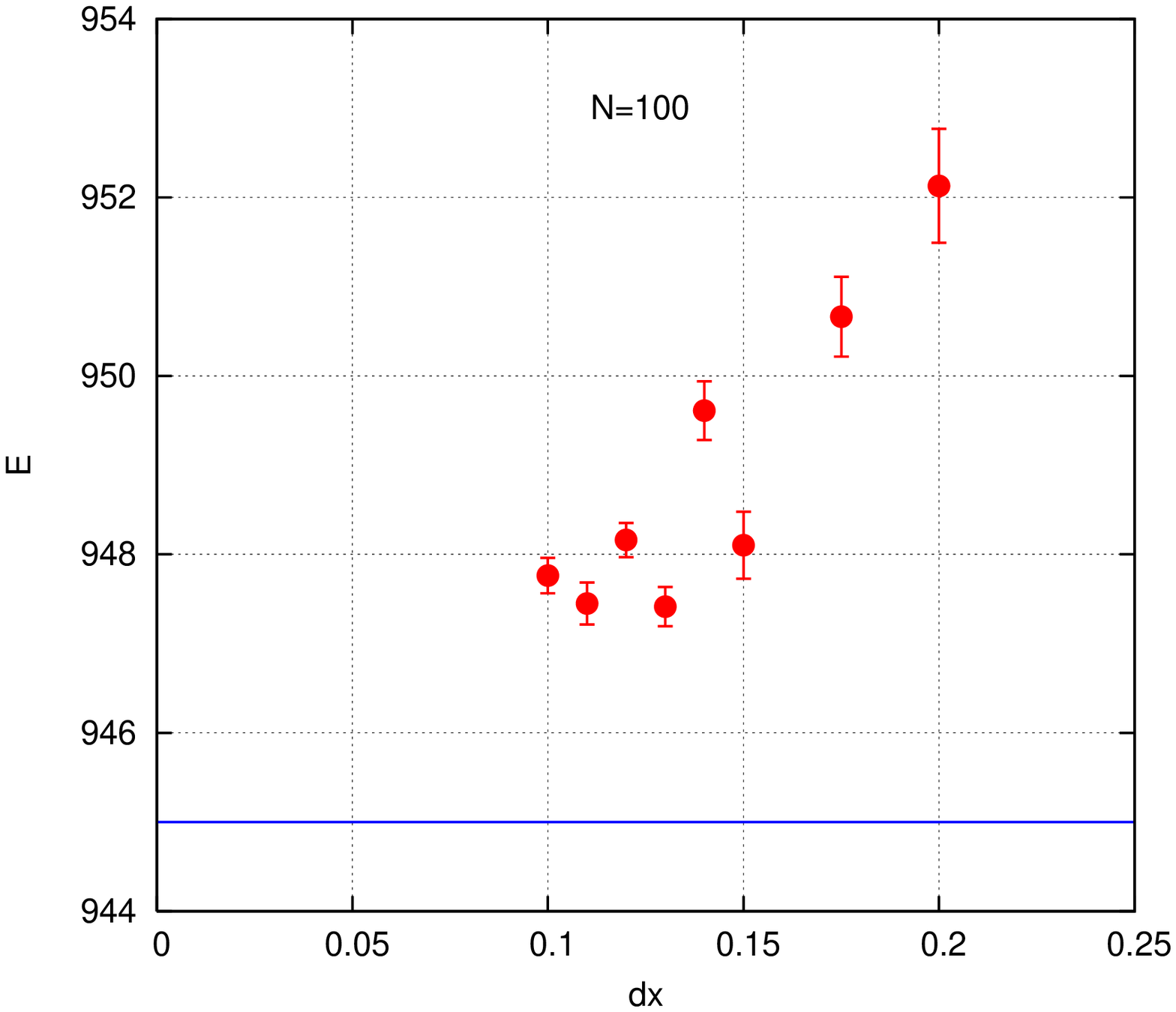}
\caption{ (color online)
 The energy of $N$ non-interacting, spin-polarized fermions in a 2D harmonic oscillator. 
For $N=40, 60, 80, 100$, the exact energies, indicated by the horizontal blue line, 
are 240, 440, 676 and 945 (in unit of $\hbar\omega$) respectively. 
The calculated lowest energies, at or near the smallest value of $\Delta x$, are 
240.06(4), 440.05(3), 677.2(1) and 947.4(2) respectively.
}
\label{FIG:g}
\la{fig1}
\end{figure}

Consider $N$ non-interacting particles in a $D$-dimensional harmonics oscillator.
According to the above discussion, each particle's ground state wave function would be (unnormalized)
$$
\psi_0(\x_i)=\exp[-\frac1{2\tau}(\x_i-\s_i)^2]
$$
with $\tau=1$ and where $\s_i=\x(\tau\rightarrow\infty)\rightarrow 0$.
However, for our purpose of anti-symmetrization, we will only let each $\s_i$ approaches 
close to zero, but not exactly zero. 
For $N$ spin-polarized fermions,
as long as all $\s_i$ are distinct, one can construction the anti-symmetric determinant wave function
\be
\Psi(\x_1,\x_2\dots\x_N)=\det\Bigl|\exp[-\frac1{2}(\x_i-\s_j)^2]\Bigr|.
\la{detwf}
\ee
Remarkably, this simple wave function gives the {\it exact} energy of $N$ non-interacting 
fermions in a harmonic oscillator so long as all $\s_i$ are close to zero but remain distinct from one another.
For the case of $D=2$, this is shown in Fig.\ref{fig1}.

These four calculations were done
by generating $N$ positions of $\s_i$ randomly near the origin with approximate
separations of $\Delta x$. This is necessary to prevent $\s_i$ from overlapping, causing the
determinant to vanish. The square of this wave function (no sign problem) is
then sampled using the Metropolis {\it et al.}\cite{met} algorithm. 
To compute the energy, it is necessary to
compute the inverse of the matrix in (\ref{detwf}).
With decreasing $\Delta x$, particles are closer to each other and
closer to the origin. For $N$ up to 60, one sees that the calculation gives the correct energy
up to statistical uncertainties. For $N>60$, there is a systematic bias due the limitation of
double precision in Fortran. When $N$ is large, the determinant is nearly vanishing and the routine 
for inverting the matrix is increasingly inaccurate. This prevents the calculation from being done 
at a $\Delta x$ sufficiently small to give the correct result. This is shown in the case of $N=80$ 
and $N=100$. The use of multiple-precision arithematics would alleviate this purely numerical problem.

This wave function (\ref{detwf}) for computing the non-interacting fermion energy is much 
simpler than anti-symmetrizing excited states of the harmonic oscillator, or using the exact 
harmonic oscillator propagator\cite{chin15}. The reason why this wave function (\ref{detwf}) is exact
can be seen from formulas given in Ref.\cite{mik01,yan02}. Here, we can give a simple example 
to illustrate the idea. For $N=2$, the (unnormalize) antisymmetrized wave function is
\ba
\Psi(\x_1,\x_2)&=&\e^{-\frac1{2}[(\x_1-\s_1)^2+(\x_2-\s_2)^2]}
-\e^{-\frac1{2}[(\x_1-\s_2)^2+(\x_2-\s_1)^2]}\nn\\
&=&\e^{-\frac1{2}[(\x_1-\s_1)^2+(\x_2-\s_2)^2]}
(1-\e^{-[(\s_1-\s_2)\cdot(\x_1-\x_2)]}).
\ea
In the limit of $\s_i\rightarrow 0$, the wave function to first-order in $\s_1,\s_2$ is just
\be
\Psi(\x_1,\x_2)=(\s_1-\s_2)\cdot(\x_1-\x_2)e^{-\frac1{2\tau}(\x_1^2+\x_2^2)},
\ee
which is proportional to the correct two fermion wave function in the harmonic oscillator. 
Note that we must have $\s_1\ne\s_2$, otherwise the wave function vanishes.

\section{Spontaneous symmetry-breaking wave functions}
\la{sbwf}

From this point onward, we will only discuss the case of $D=2$.
For $N$ fermions in a harmonic oscillator with Coulomb interactions, the Hamiltonian
is given by\cite{gho07}
\be
H=-\frac12\sum_{i=1}^N\nabla_i^2+\frac12 \sum_{i=1}^N\x_i^2+\sum_{i>j}\frac{\ld}{x_{ij}}.
\la{mbh}
\ee
where $x_{ij}=|\x_i-\x_j|$.
The similarily transformed propagator will yield the corresponding anti-symmetric wave function
\be
\Psi_D(\x_1,\x_2\dots\x_N)=\det\Bigl|\exp[-\frac1{2\tau}(\x_i-\s_j)^2]\Bigr|.
\la{detwfn}
\ee
Here, we will let the variance of the Gaussian distribution, $\tau$, usually set to 1, be allowed to vary.
As before,  each $\s_i=\x_i(\tau\rightarrow\infty)$ is a {\it stationary} point of the
trajectory $\x_i(\tau)$ obeying the drift equation
\be
\frac{d\x_i}{d\tau}=-\nabla_i S(\x_1,\x_2\dots,\x_N),
\la{dfeq}
\ee
with $S(\x_1,\x_2\dots,\x_N)$ being the action of the 
many-particle {\it bosonic} ground state wave function:
$$
\Psi_B(\x_1,\x_2\dots\x_N)=\e^{-S(\x_1,\x_2\dots,\x_N)}.
$$
Note that the set of stationary points satisfying $d\x_i/d\tau=0$ correspond to
$\nabla_i S(\{\x_i\})=0$, and are positions which minimize $S(\{\x_i\})$,
or {\it maximize} the bosonic wave function. (The case of multiple local maxima will
be discuss in later Sections.) In the non-interacting case, we have seen in the previous section 
that anti-symmetrizing the exact bosonic ground state produces the exact fermionic ground state. 

With the added Coulomb interaction, the exact bosonic ground state is known only for two particles 
at coupling $\ld=1$ with 
\be
S(\x_1,\dots \x_N)=\frac12 \sum_{i=1}^2\x^2_i-\ln(1+x_{12}),
\la{n2ex}
\ee
and $E_0=3$. The drift equations from (\ref{dfeq}) are then
$$
\frac{d\x_1}{d\tau}=-\x_1+\frac{\hat\x_{12}}{1+x_{12}},
\qquad \frac{d\x_2}{d\tau}=-\x_2+\frac{\hat\x_{21}}{1+ x_{12}},
$$
$$
\rightarrow\quad \frac{d\x_{cm}}{d\tau}=-\x_{cm},
\quad{\rm and}\quad  \frac{d\x_{12}}{d\tau}=-\x_{12}+\frac{2\hat\x_{12}}{1+x_{12}}.
$$
Since the drift equations are just first-order differential equations, they can be solved easily
by any numerical method to arrive at their statinary points. In the above case, the stationary 
points can be gotten simply by setting the $\tau$-derivatives to zero:
$$\s_{cm}=0\quad {\rm and}\quad \s_{12}=\hat\x_{12}(0)\quad\rightarrow
\quad \s_{1}=\frac12 \hat\x_{12}(0),\quad \s_{2}=-\frac12 \hat\x_{12}(0).
$$
The two stationary points $\s_1$ and $\s_2$ are antipodal points on a circle of radius $R=1/2$, 
oriented by the initial vector $\hat\x_{12}(0)$, which is entirely arbitrary. 
Thus any two such antipodal points on the circle can be stationary points of the 
above drift equation. However, when a specific pair of points 
is inserted into the fermion wave function (\ref{detwfn}), the resulting wave function no longer 
respects the rotational symmetry of the original Hamiltonian. Thus the transformed propagator 
naturally produces a {\it spontaneous symmetry-breaking wave function}, which has been 
extensively discussed in the literature\cite{yan99,mik01, yan02}, notably by Yannouleas and Landman\cite{yan99,yan02,yan07}. 
In these earlier discussions, such a wave function 
was simply viewed as an ansatz, and it is therefore entirely reasonable to take $\{\s_i\}$
as particle positions which minimize the classical potential energy\cite{mik01,yan02,kai02}. In this case, they would be
antipodal points on a circle with $R=2^{1/3}/2=0.62996$. Here, our derivation of this wave function 
from the transformed propagator showed that {\it these stationary points are to be determined 
by the {\it maximum} of the bosonic wave function}. In Fig.\ref{n23} we compare the 
energies computed from the fermion wave function (\ref{detwfn}) using these two sets of stationary 
points with that from a
5-bead fermion PIMC calculation using an optimized fourth-order propagator, as described in Ref.\cite{chin15}. 
The top line gives the energy from using stationary points minimizing the potential energy. 
The bottom line gives the energy from using stationary points maximizing 
the bosonic wave function. This comparison clearly shows that one should use stationary points from the
latter rather than from the former. Moreover, 
the fermion wave function (\ref{detwfn}) is optimal with $R=1/2$; any other radius yields a higher energy.

\begin{figure}[hbt]
\includegraphics[width=0.49\linewidth]{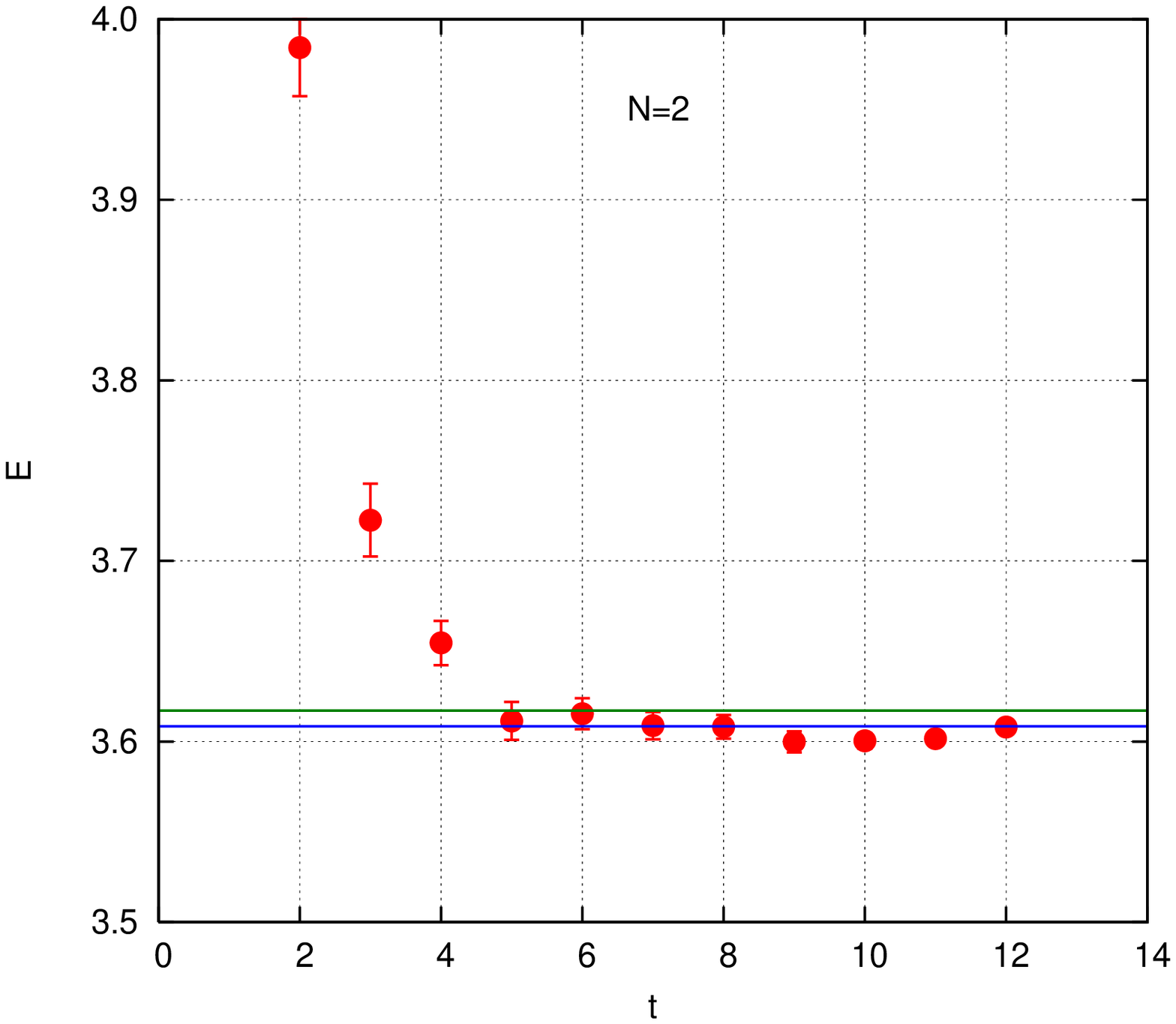}
\includegraphics[width=0.49\linewidth]{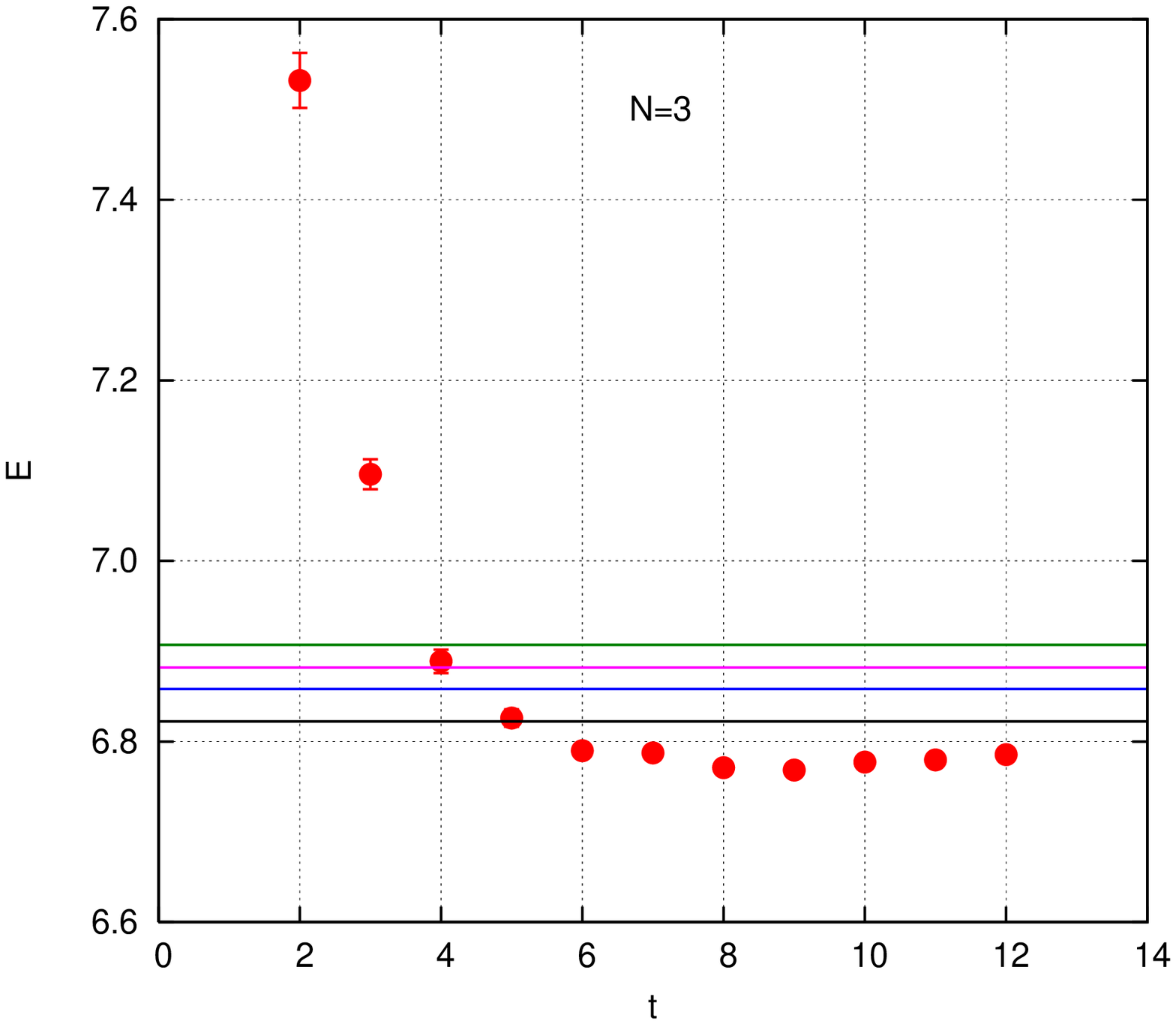}
\caption{ (color online)
{\bf Left}: Two-fermion energies at coupling $\ld=1$.
Symbols are results from a 5-bead PIMC calculation using an optimized fourth-order fermion propagator\cite{chin15}
yielding a minimum energy of 3.600(4) at imaginary time $t=10$. 
The top and the bottom line denote energy 3.6171(6) and 3.6085(5) respectively. See text for detail.
{\bf Right}: Three-fermion energies at coupling $\ld=1$. Symbols
show the 5-bead PIMC energy of 6.768(4) at $t=9$. From the top down, the four horizontal lines are
three-fermion energies of 6.907(1), 6.882(1), 6.858(1), 6.822(1) respectively, computed from various forms of the wave function (\ref{detwfn}). See text for detail. 
}
\la{n23}
\end{figure}

The rotational symmetry of this wave function can be restored by integrating over the 
angle of $\hat\x_{12}(0)$, bascially summing the wave function over all antipodal points on the circle.
Such a {\it symmetry-restored} wave function\cite{yan02,yan07} 
should have lower energy and {\it may} account for
the difference of $0.0085(5)$ between this wave function's energy and that of PIMC.
In this work, we will not pursue this symmetry-restoration energy correction. 

For three particles, the exact bosonic ground state is unknown. However,
from the above discussion, by symmetry, the three stationary points must form
an equilateral triangle with energy minimized by their distance from the origin. 
To control the overall size of the triangle, we do not need the exact bosonic ground state;
it is sufficient to use a trial ground state with action
\be
S(\x_1,\x_2\dots,\x_N)=\frac12\sum_{i=1}^N\x_i^2-\sum_{i>j}\frac{\ld x_{ij}}{1+b x_{ij}}.
\la{bs}
\ee
Here, the pairwise correlation function is well known to satisfy the 2D
cusp condition with parameter $b$ varying the strength of the correlation. (The cusp
condition here is due to the bosonic ground state only, and has nothing to do with
whether the two particles are in a relative a spin-triplet or singlet state.) The resulting
drift equation 
\be
\frac{d\x_i}{d\tau}=-\nabla_i S(\x_1,\x_2\dots,\x_N)
=-\x_i+\sum_{j\ne i}\frac{\ld\hat\x_{ij}}{(1+b x_{ij})^2},
\la{deq}
\ee
can be solved numerically for any $N$ to obtain the set of stationary points $\{\s_i\}$.
With this correlator, as $\ld\rightarrow 0$, $\s_i\rightarrow 0$, and
the wave function (\ref{detwfn}) reduce to the exact wave function for
$N$ non-interacting fermions of the last Section.

At the right of Fig.\ref{n23}, we compare the three-fermion energy at coupling $\ld=1$ using
various form of the wave function (\ref{detwfn}) to that of a 5-bead PIMC calculation.
The top most horizontal line is the energy resulting from using stationary points from minimizing 
the potential energy. The equilateral triangle is at $R=0.83$. The next line down uses the 
correlation function of the exact two-fermion solution (\ref{n2ex}) giving $R=0.75$. 
This shows that the correlation function which is exact for two-body may not
be good enough for more than two bodies. 
The third line gives the energy using (\ref{bs}) at $b\approx 1$, but keeping $\tau=1$, 
yielding $R\approx 0.50$.
Finally, the lowest line corresponds to allowing $\tau$ to vary in additional to $b$.
The minimum energy at $b=1.7$, $\tau=1.1$, which shrank $R$ to $\approx 0.38$ but 
broadened the Gaussian, is substantially better than varying $b$ alone. 
The resulting energy is above the PIMC result by less than one percent. 

As shown in Section\ref{non}, since our determinant wave function is
exact in the non-interacting limit, it should be good at weaking couplings. 
We therefore test the wave function here in the strong coupling limit of $\ld=8$.
In Table \ref{tab1}, the resulting energies from this two-parameter wave function 
for a 2D quantum dots with up $N=100$ spin-polarized electrons are shown under the column SBWF,
short for ``Symmetry-Breaking Wave Function". The SBWF energies at this strong coupling 
 are comparable to the 2-bead, fourth-order propagator results B2 from Ref.\onlinecite{chin15}.  
Since B2 is still a PIMC calculation, the energy needs to be extracted at an imaginary 
time of $\tau\approx 3-4$.  At this value of $\tau$, with more particles, the free fermion determinant 
propagator is increasingly near zero, and its inversion needed for computing the Hamiltonian estimator limits 
the particle size to $N\approx 40$. Here, SBWF is like that of a free determinant propagator
at only $\tau=0.4-0.8$ and therefore can be used for up to $N=100$ fermions, or more.
Energies in other columns will be described in the next Section.

\begin{table}
\caption{Comparison of $N$ {\it spin-polarized} 2D electron ground state energies 
$E_0/\hbar\omega$ at coupling $\lambda=8$.}
\begin{center}
\begin{tabular}{c r r r r r r r r r}
\colrule
$N$ &  $\tau\ $ &{\rm b}\ \ &\ \ \ \ {\rm SBWF}\ \ \  &\  {\rm B2\cite{chin15}}\ \ & GSPI2\ \ \ & GSPI4\ \ \ & 
PIMC\cite{egg99}& CI\cite{ron06}\ &\ \ DMC\cite{ped03,gho07} \\
\colrule
4     &0.80&0.60   &      28.217(3)       &     28.266(5) &27.927(3)& 27.818(5)& 27.823(11) &27.828   &     \\
6     &0.80&0.65   &      61.257(5)       &     61.403(7) &60.686(4)&60.475(6) & 60.42(2)&60.80 & 60.3924(2)      \\
8     &0.70&0.67   &     104.21(1)        &    104.45(1) &103.425(8)&103.161(9) &  103.26(5)&&103.0464(4)       \\
\colrule
10   &0.70&0.68   &     156.75(1)         &   156.77(1) & 155.57(1)&155.23(1)&         \\
20   &0.65&0.70   &     537.56(2)         &   538.07(3) &534.71(5)&534.1(1) &         \\
30   &0.60&0.75   &   1091.60(4)         &  1091.7(1)&1086.5(1) &1085(1)&         \\
40   &0.60&0.74   &   1795.74(9)         &   1795.9(1) & 1787.9(5)&         \\
50   &0.55&0.76   &   2636.73(6)        &   &2627.0(3) &         \\
60   &0.50&0.78   &   3604.45(7)\        &   &3593(1)\ \ \ &         \\
80   &0.50&0.78   &   5893.2(3)\ \      &   & &         \\
100 &0.45&0.80   &   8618.1(3)\ \      &   & &         \\
\colrule
\end{tabular}
\end{center}
\label{tab1}
\end{table}

With increasing number of fermions, Table \ref{tab1} shows that $b$ increases, weakening the 
inter-particle repulsion, and $\tau$ decreases, making each Gaussian smaller. 
Both act to increase the particle density, but the quantum dot continues to expand in size with 
increasing number of fermions. This is shown in Fig.{\ref{pos200}, where
the stationary points of wave function (\ref{deq}) is plotted for 10, 30, 60 and 100 particles, with 
dot radius set equal to $\sqrt\tau$. (This gives a crude picture of the one-body density of the 
Bosonic wave function.)
While the stationary points' concentric ring-like structure is very clear for
10 to 60 particles, and is similar to those determined by the classical potential energy\cite{bed94}, 
this ring-like structure is less clear for 100 particles.
With increasing number of particles, there are many stationary 
configurations which are not strictly ring-like and only differ minutely in energy.
Our algorithm for solving the drift equation (\ref{deq}) simply
evolves a set of random initial positions for a long time, and therefore has no way of picking
out only concentric ring-like configurations.  It is also possible that at large $N$,
rotational symmetry is broken entirely without any trace of discrete circular symmetry.

\begin{figure}[hbt]
\includegraphics[width=0.49\linewidth]{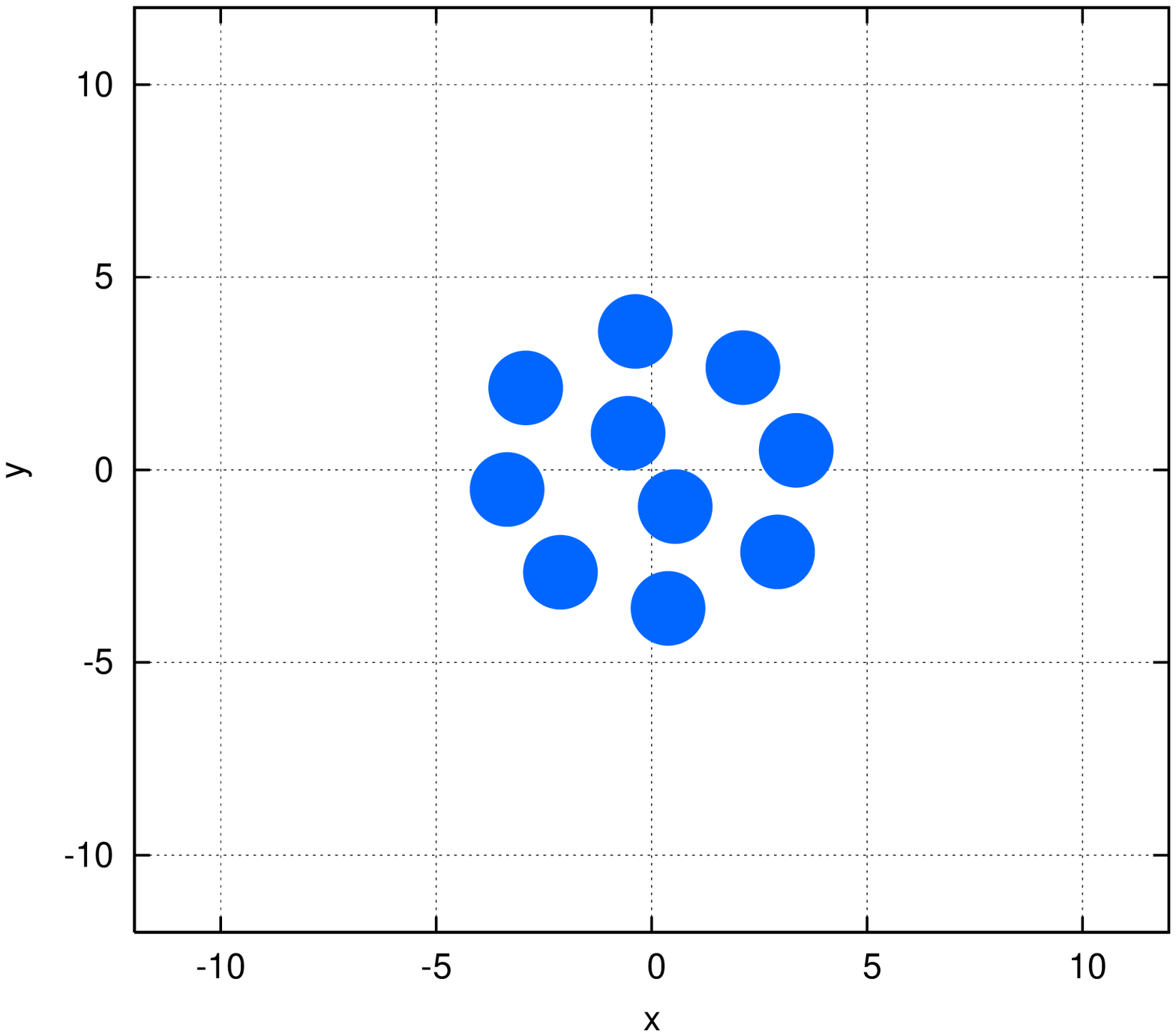}
\includegraphics[width=0.49\linewidth]{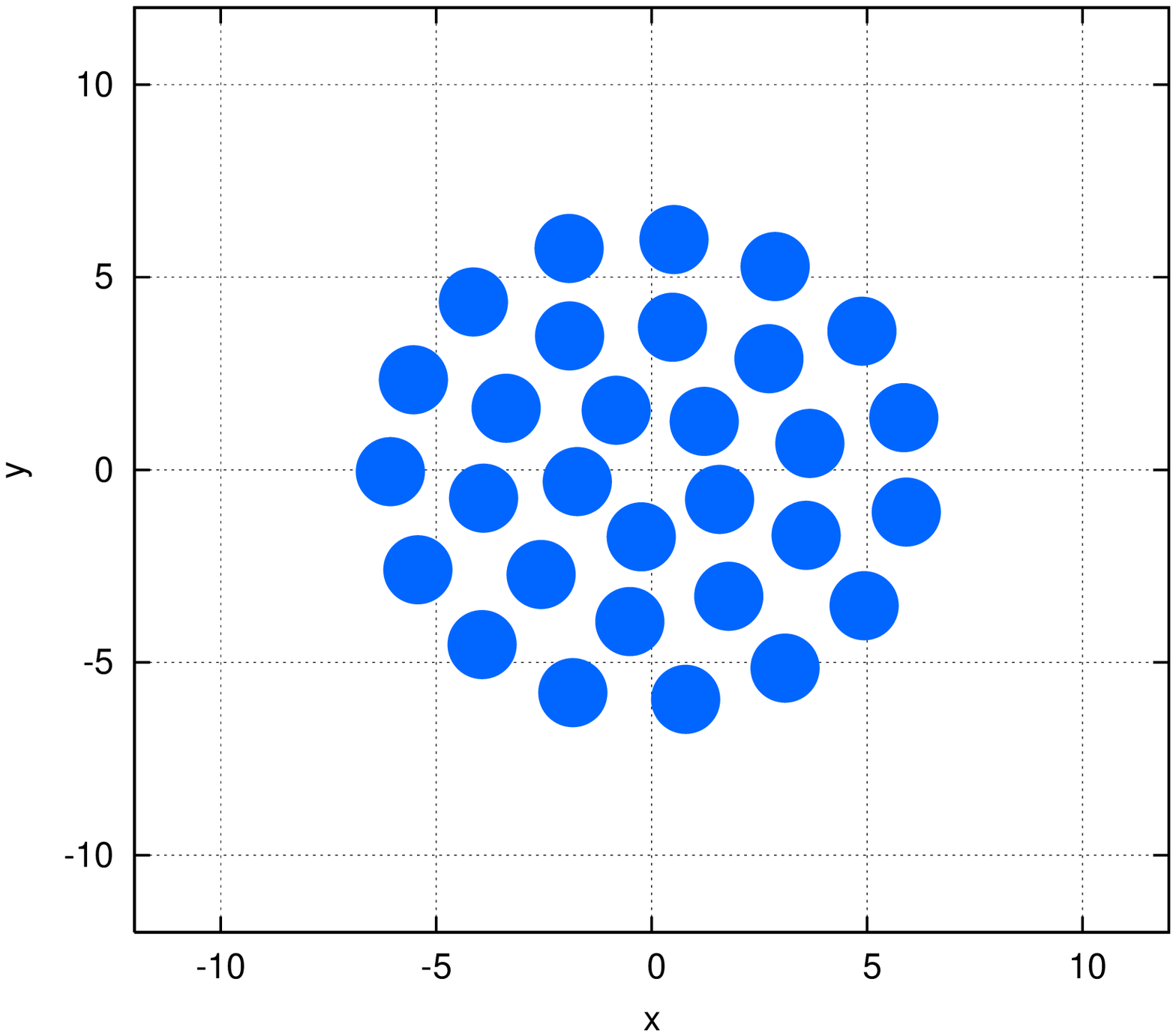}

\includegraphics[width=0.49\linewidth]{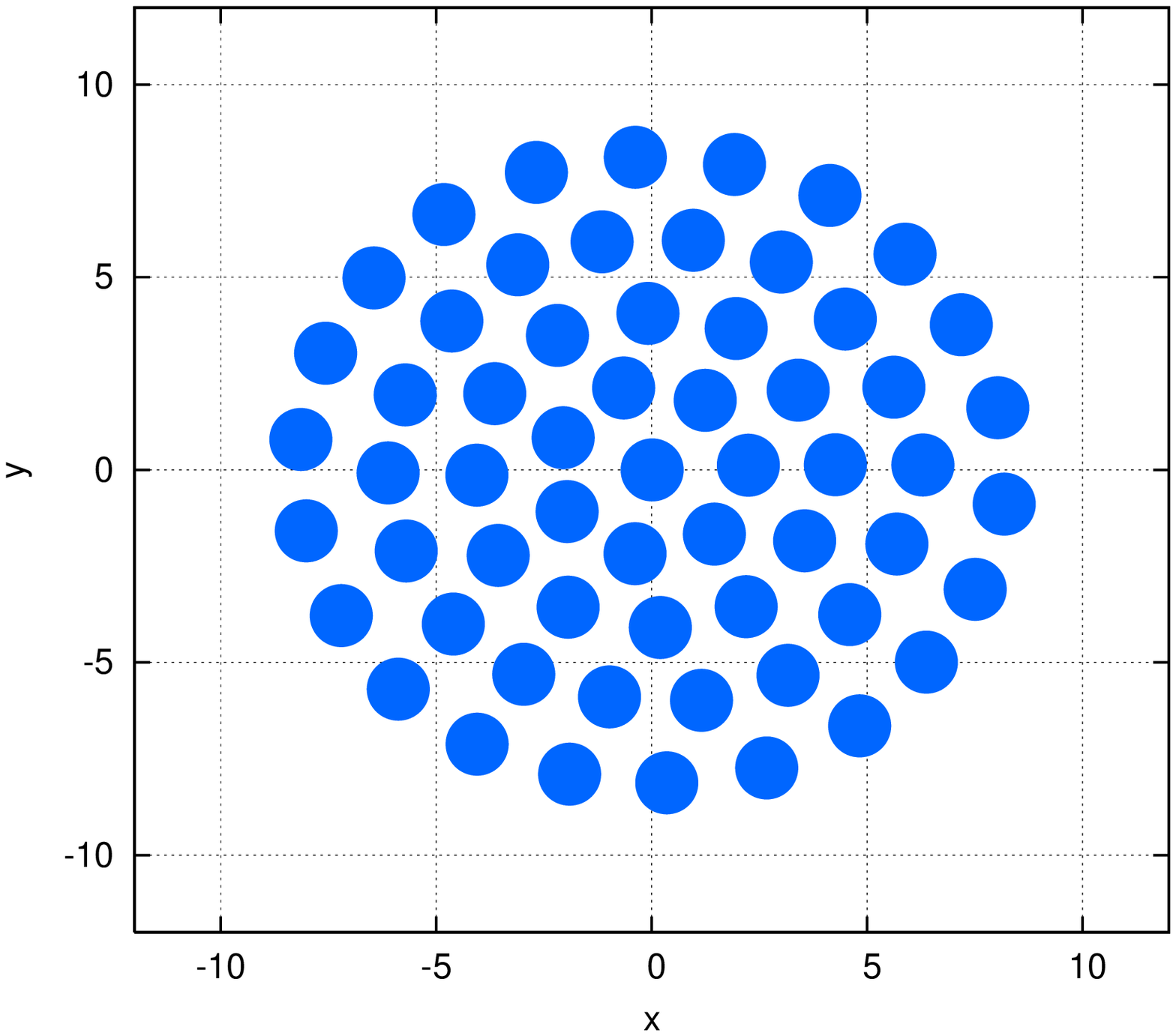}
\includegraphics[width=0.49\linewidth]{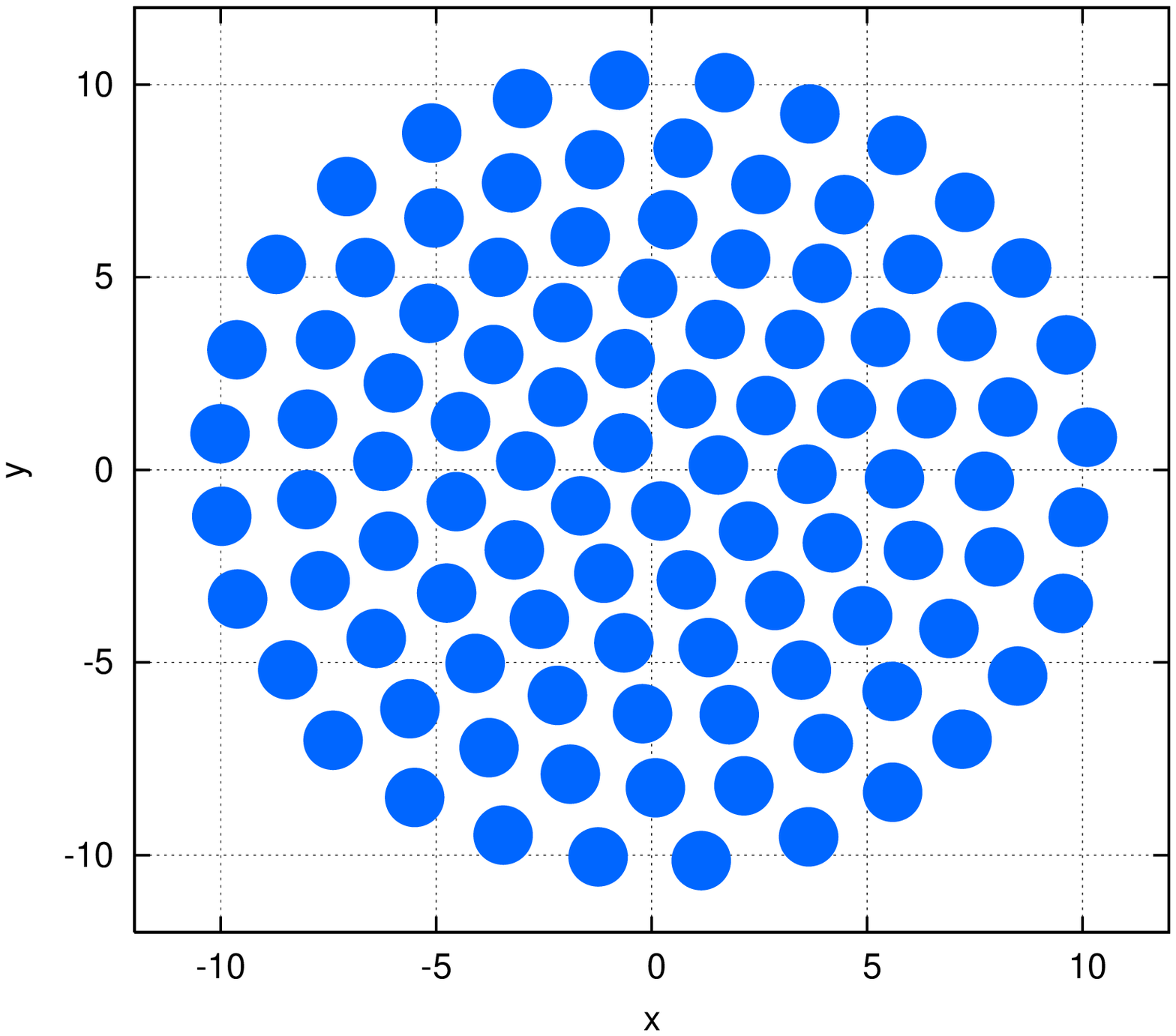}
\caption{ (color online)
Stationary points for the 10, 30, 60 and 100-particle wave function at $\ld=8$
with each dot's radius set equal to $\sqrt\tau$.
}
\la{pos200}
\end{figure}

\section{Fermion Ground State PIMC}
\la{gspimc}

   As we have shown in the last section, the determinant wave function (\ref{detwfn}) allows 
one to obtain excellent variational energies for up to 100 fermions (or spin-polarized electrons) 
in a 2D quantum dot. 
To lower the energy further, one can do a Fermion Ground State Path Integral Monte Carlo (FGSPIMC) 
calculation based on that trial function via

\be
E_0=\lim_{\tau\rightarrow\infty}\frac{\int d\bx' d\bx_1d\bx \Psi_D(\bx')G(\bx',\bx_1;\tau)H G(\bx_1,\bx;\tau)\Psi_D(\bx)}
{\int d\bx' d\bx_1d\bx\Psi_D(\bx')G(\bx',\bx_1;\tau)G(\bx_1,\bx;\tau)\Psi_D(\bx)},
\la{gspi}
\ee
where $G(\bx',\bx;\tau)$ can be either the commonly used second-order primitive propagator
$$
G_2(\bx',\bx;\tau)=\e^{-\frac{\tau}{2}V(\bx')}\det\Bigl|\exp[-\frac1{2\tau}(\x'_i-\x_j)^2]\Bigr|
\e^{-\frac{\tau}{2}V(\bx)},
$$
or the fourth-order propagator corresponding to B2 of Ref.\onlinecite{chin15}. To preserve the upper bound property of the Hamiltonian
estimator, it is necessary to evaluate $H$ only at the middle of the integral. With anti-symmetric 
propagators, evaluating $H$ at any other position destroys this upper bound
property. This greatly limited the choice of $G(\bx',\bx;\tau)$. If $G_2(\bx',\bx;\tau)$ is used,
then (\ref{gspi}) is a four-bead calculation, having essentially four anti-symmetric free-propagators.
If the fourth-order propagator $G_4(\bx',\bx;\tau)$ is used, each requiring two 
anti-symmetric free-propagators, then (\ref{gspi})
 is a six-bead calculation. Both will then have sign problems, with the latter more servere. However,
this GSPIMC calcuation will still be better than doing a straightforward PIMC calculation. This is because
for a PIMC calculation, the ground state energy can only be extract at a relatively large imaginary time, 
such as $\tau\approx 8$, whereas evolving from $\Psi_D(\bx)$, one only needs $\tau\approx 3$ or less. 
This then greatly reduces the sign problem for determining the ground state energy of a
large quantum dot.

\begin{figure}[hbt]
\includegraphics[width=0.95\linewidth]{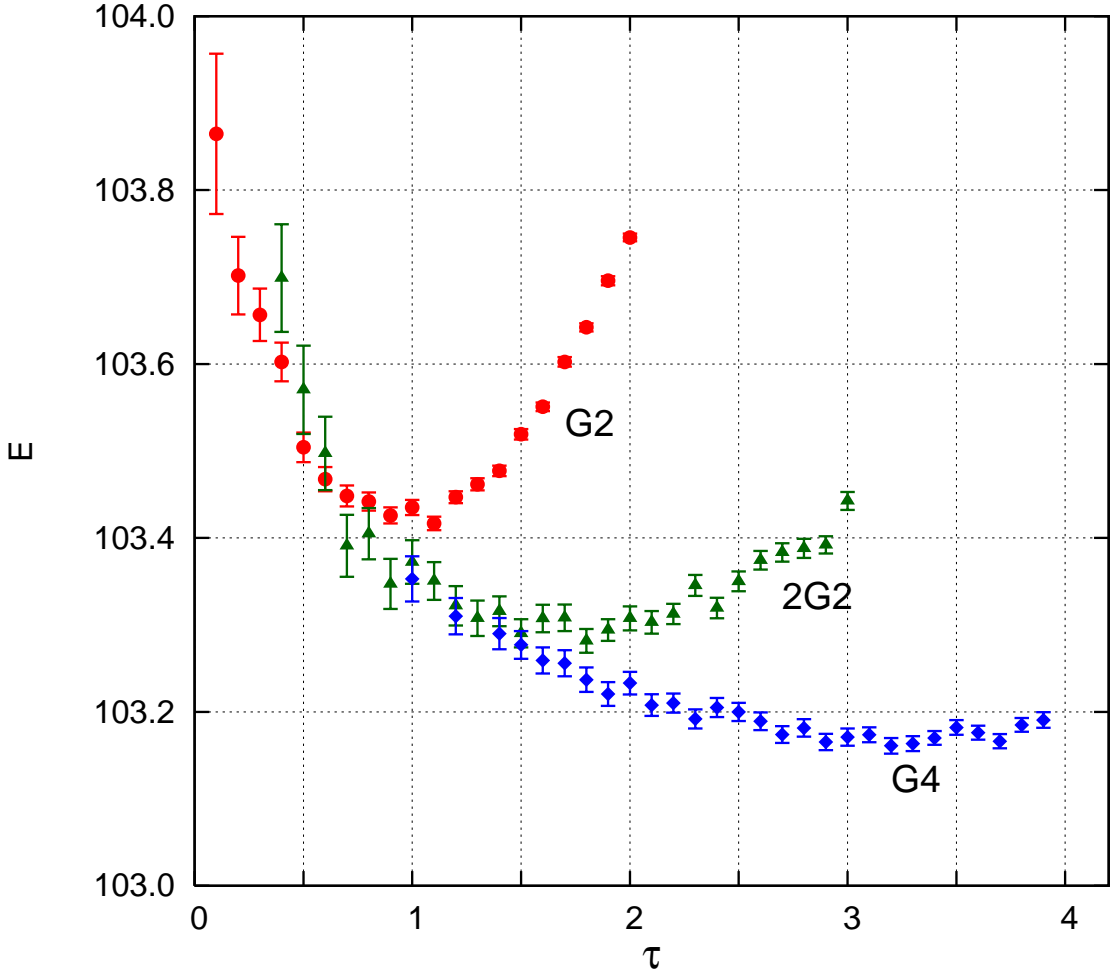}
\caption{ (color online)
Ground state PIMC calculations for 8 spin-polarized electrons in a 2D harmonic oscillator at coupling $\ld=8$
via (\ref{gspi}). $G_2$ and $G_4$ are second and fourth-order imaginary time propagators respectively.
}
\la{fign8}
\end{figure}

In Fig.\ref{fign8}, we show various GSPIMC calculations for the ground state energy of 8 spin-polarized
electrons at $\ld=8$. Since $G_4$ uses two free-fermion propagators, we also computed the case with 
$2G2(\tau)=G_2(\tau/2)G_2(\tau/2)$, which is two second-order propagators at half the time step. 
The dramatic improvement in using $G_4$ is clearly visible. 
The flatness of the energy curve at large $\tau$ argues strongly that its energy is close to being exact.
This is indeed the case, as shown in Table \ref{tab1}, where the single $G_2$ and $G_4$ energies
are shown under columns GSPI2 and GSPI4 respectively. While $G_4$ clearly refines the energy
toward the exact, its improvement over that of $G_2$ is a mere $\approx 0.2\%$ in the case
of $N$=8. By comparison, $G_2$ lowers the SBWF energy by $\approx 0.8\%$.  Since $G_4$  is
a six-bead calculation, due to the sign problem, it can only be used up to $N\approx 30$. $G_2$
remains effective for quantum dots twice as large, up to $N\approx 60$.

While the use of GSPIMC for solving bosonic systems is fairly common, its application to fermions,
due to the sign problem, has not been as prevalent. Ref.\onlinecite{cal14} considered various choices for the free 
propagator, but if one views (\ref{gspi}) as an extension of fermion PIMC, then the natural
choice is to use of the anti-symmetric determinant propagator.

\section{Conclusions and future directions}
\la{con}

   In this work, we have shown that 1) similarity-transformed propagators, in the case of quantum dots, 
can naturally produce spontaneous symmetry-breaking (SSB) wave functions for solving many-fermion problems. 
This is a theoretical advance in that such a SSB wave function was previously regarded only as
an {\it ad hoc} ansatz. 2) Our derivation show that the particle positions of such a SSB wave function
should be determined by maximizing the many-body bosonic wave function, rather than just
minimizing the potential energy. 3) The use of such SSB wave function in solving the many-fermion problem via VMC
is far simpler than using a determinant of excited states plus Jastrow correlators. 4) We have further demonstrated
the usefulness of using higher order propagators in the context of doing fermion GSPIMC.

   A natural generalization of this work is to solve for case of spin-balanced quantum dots\cite{ilk17}, with equal
 number of spin-up and spin-down electrons. However, the resulting SSB wave function now works less well due
 to {\it spin-frustration}. Take the example of $N=6$ with $N_{\uparrow}=3$ and $N_{\downarrow}=3$. 
 In each case of $N_{\uparrow}=3$ or $N_{\downarrow}=3$, the preferred configuration is an equilateral triangle.
 Therefore, for $N=6$, the minimum energy configuration should be the interlacing of two equilateral triangles,
forming a hexagon, with alternating spin at each vertex. However, for $N=6$, the configuration which maximizes the
bosonic wave function (or that of minimizing the potential energy) is a pentagon with a single particle at the center\cite{bed94}. Therefore one spin-up (or down) particle must be at the center. Such a wave function then frustrates the desired spin assignment and further breaks the spin-up/spin-down symmetry of the system. 
Moreover, for $N_{\uparrow}=N_{\downarrow}=N/2$, there are $N!/(2(N/2)!)$ distinct ways of assigning $N/2$ up spins and all are possible SSB wave functions. At this time, there is no known rule for determining which spin state will give the lowest energy. Alternatively, one can try to restore the spin-symmetry by summing over all states of distinct spin configurations. Such a multi-determinant calculation would require an order of magnitude more effort and would be best done in a future publication.

  For atomic calculations, since the Hatree-Fock method generally works well and gives no indication of any SSB state, the method proposed here will probably not be applicable. However, for nuclei calculations, since our method is exact for non-interacting fermions in a harmonic oscillator (which is the basis of the shell-model), our method may be useful for calculating alpha-particle clustered nuclei such as $C^{12},O^{16},N\!e^{20},$ etc., since alpha-particle clustering  can be viewed as a SSB state.

\begin{acknowledgments}
Portions of this research were conducted with the advanced computing resources provided by Texas A\&M High Performance Research Computing.
\end{acknowledgments}


\newpage
\begin{thebibliography}{99}

 \bibitem{rei02} S. M. Reimann and M. Manninen, Rev. Mod. Phys. 74, 1283
(2002).

 \bibitem{bli95} S. M. Blinder, J. Math. Phys. 36, 1208 (1995)
 
\bibitem{nei62}N. R. Kestner and O. Sinanoglu,
Phys. Rev. 128, 2687 (1962) 
 
 \bibitem{yan99} C. Yannouleas and U. Landman
 Phys. Rev. Lett. 82, 5325 (1999)– Erratum Phys. Rev. Lett. 85, 2220 (2000)

 \bibitem{fer94} M. Ferconi and G. Vignale
Phys. Rev. B 50, 14722 (1999)

 \bibitem{hir99} K. Hirose and N. S. Wingreen
Phys. Rev. B 59, 4604 (1999).

\bibitem{ron06} M. Rontani, C. Cavazzoni, D. Bellucci and G. Goldoni
 J. Chem. Phys. {\bf 124}, 124102 (2006).

\bibitem{wal} E. Waltersson, C. J. Wesslen, and E. Lindroth,
Phys. Rev. B 87, 035112 (2011).

\bibitem{ped} M. Pedersen Lohne, G. Hagen, M. Hjorth-Jensen, S. Kvaal, and F. Pederiva,
Phys. Rev. B 84, 115302 (2011).

\bibitem{har02}
A. Harju, S. Siljamaki, and R. M. Nieminen
Phys. Rev. B 65, 075309 (2002)

\bibitem{kai02} J. Kainz, S. A. Mikhailov, A. Wensauer, and U. R\"ossler,
Phys. Rev. B 65, 115305 (2002).

\bibitem{ped03}F. Pederiva, C.J. Umrigar and E. Lipparini, 
Phys. Rev. {\bf B 62}, 8120 (2000), {B 68}, 089901(E) (2003)

\bibitem{gho07}A. Ghosal, A. D. G\"ucl\"u, C. J. Umrigar, D. Ullmo and H. U. Baranger 
Phys. Rev. {\bf B 76}, 085341 (2007)

\bibitem{mak98} C. H. Mak, R. Egger, and H. Weber-Gottschick, Phys. Rev. Lett. {81}, 4533 (1998).

\bibitem{egg99}R. Egger, W. H\"ausler, C. H. Mak and H. Grabert, Phys. Rev. Lett. {82}, 3320 (1999);
{\bf 83}, 462(E) (1999). 

\bibitem{egg00}  R. Egger, L. M\"uhlbacher, and C. H. Mak, Phys. Rev. {E 61}, 5961 (2000).

\bibitem{reu03}  B. Reusch and R. Egger, Europhys. Lett. 64, 84 (2003).

\bibitem{chin15} Siu A. Chin, Phys. Rev. E 91, 031301(R) (2015).

\bibitem{ilk17} Ilkka Kylanpaa and Esa Rasanen
Phys. Rev. B 96, 205445 (2017)

\bibitem{cep95}
  D.~M. Ceperley,
  Rev. Mod. Phys., {\bf 67}, 279 (1995).

 \bibitem{sar00} A. Sarsa, K. E. Schmidt, and W. R. Magro, 
J. Chem. Phys. 113, 1366 (2000).

\bibitem{cal14} F. Calcavecchia, F. Pederiva, M. H. Kalos, and
Thomas D. K\"uhne,
Phys. Rev. E 90, 053304 (2014) 

\bibitem{mik01} S. A. Mikhailov,
Physica B 299, 6 (2001).

\bibitem{yan02} C. Yannouleas and U. Landman,
Phys. Rev. B 66, 115315 (2002).


\bibitem{yan07} C. Yannouleas and U. Landman,
Rep. Prog. Phys. 70 (2007) 2067–2148.



\bibitem{chin90} Siu A. Chin, ``Quadratic diffusion Monte Carlo algorithms for solving atomic
many-body problems'', Phys. Rev. A 42, 6991 (1990).

\bibitem{mos82} J. W. Moskowitz, K. E. Schmidt, M. E. Lee, and M. H. Kalos,
J. Chem. Phys. 77, 349 (1982).

\bibitem{rey82}P. J. Reynolds, D. M. Ceperley, B. J. Adler, and W. A. Lester,
J. Chem. Phys. 77, 5593 (1982).

\bibitem{caf88} M. Caffarel and P. Claverie,
J. Chem. Phys. 88, 1088 (1988); 88, 1100 (1988).

\bibitem{uhl} G. E. Uhlenbeck and L. S. Ornstein, 
 Phys. Rev. 36,  823–841 (1930)
 
 \bibitem{met} N. Metropolis, A. W. Rosenbluth, M. N. Rosenbluth, A. H. Teller and E. Teller, 
 J. Chem. Phys. 21, 1087 (1953)
 
 \bibitem{bed94}V. M. Bedanov and F. M. Peeters
  Phys. Rev. B 49,  2667–2675 (1994)




\end{thebibliography}
\end{document}